\documentclass[twocolumn,noshowpacs,prl,preprintnumbers,superscriptaddress,amsmath,amssymb,floatfix]{revtex4}
\usepackage{graphicx}
\usepackage{subfiles}
\usepackage{placeins}
\usepackage{hyperref}
\usepackage{pdfpages}
\usepackage{mwe,tikz,fix-cm}
\usepackage{titlesec}
\usepackage[caption=false]{subfig}

\begin{document}
	
	\title{Faster calculation of the percolation correlation length on spatial networks}
	\author{Michael M. Danziger}
	\affiliation{Center for Complex Network Research, Northeastern University, Boston, MA 02115, USA}
	\author{Bnaya Gross}
	\affiliation{Department of Physics, Bar Ilan University, Ramat Gan, Israel}
	\author{Sergey V. Buldyrev}
	\affiliation{Department of Physics, Yeshiva University, New York, New York 10033, USA}
	\date{\today}
\begin{abstract}
The divergence of the correlation length $\xi$ at criticality is an important phenomenon of percolation in two-dimensional systems.
Substantial speed-ups to the calculation of the percolation threshold and component distribution have been achieved by utilizing disjoint sets, but existing algorithms of this sort cannot measure the correlation length.
Here, we utilize the parallel axis theorem to track the correlation length as nodes are added to the system, allowing us to utilize disjoint sets to measure $\xi$ for the entire percolation process with arbitrary precision in a single sweep.
This algorithm enables direct measurement of the correlation length in lattices as well as spatial network topologies, and provides an important tool for understanding critical phenomena in spatial systems. 
\end{abstract}
	\maketitle

\section{Introduction}
Percolation describes the transition that occurs when connections are added amongst a disconnected set of nodes and global connectivity emerges~\cite{staufferaharony,bunde1991fractals}.
Originally developed to model flow in a disordered media, percolation theory has since been used to describe the robustness of complex networks~\cite{albert-nature2000,cohen-prl2000} and multilayer interdependent networks~\cite{buldyrev-nature2010,danziger-chapter2016} as well as a broad range of other phenomena.
The percolation transition is characterized by the emergence of a giant component which is of order of the system size.  
The critical exponent $\nu$ describes the divergence of the correlation length at criticality
\begin{equation}
\xi \sim (p - p_c)^{-\nu},
\end{equation}
where $p$ is the fraction of nodes in the system and $p_c$ is the critical fraction of nodes at which a giant component emerges.
Conventionally, the correlation length is calculated by traversing each cluster, and with the relative locations of each node used to calculate $R_\mu^2$, the square radius of gyration of cluster $\mu$ with $\xi$ calculated as:
\begin{equation}
 \xi^2 =\frac{ \sum_\mu m_\mu^2 R_\mu^2}{\sum_\mu m_\mu^2}
\end{equation}
where $m_\mu$ is the cluster size and the sum is over all of the clusters~\cite{staufferaharony}.
This method requires a complete sweep of the system for every value of $p$ under consideration, making a detailed measurement of $\xi$ for all values of $p$ computationally prohibitive.
The standard algorithm for calculating this is called the ``burning'' algorithm~\cite{herrmann-jphysa1984}. 

The method of disjoint sets provides substantial speed-ups for calculating the percolation threshold, cluster distribution and critical exponents.
It accomplishes this by adding nodes (or links) to the system and tracking which cluster each node belongs to, merging and relabeling clusters as necessary.
The disjoint-set data structure was first developed by computer scientists in 1964~\cite{galler-acm1964} and was applied to percolation by Newman and Ziff~\cite{Newman-Ziff-percolation-algorithm2000,Newman-Ziff-percolation-algorithm2001}, allowing for unprecedented speed and precision.
However, it is not straightforward to calculate the correlation length in this manner because upon merging clusters, the average distances change non-trivially.
For this reason, in previous percolation research the critical exponent for the correlation length ($\nu$) was measured indirectly via the wrapping probability~\cite{Newman-Ziff-percolation-algorithm2001}.
Other approaches have found speedups by exploiting the particular structure of the grid, a faster method was developed~\cite{hoshen-pre1997} but it is not adaptable to spatial disorder.

Here we introduce a new method which uses the parallel axis theorem in order to update the correlation length upon merging clusters and allows for a precise and direct measurement of the correlation length for all values of $p$ in a single network sweep.
This is important for applications such as interdependent lattices which require exact values of $\xi$~\cite{wei-prl2012,danziger-jcomnets2014,berezin-scireports2015} and particularly for spatial networks that deviate from the simple grid topology~\cite{danziger-epl2016,grossvaknin-JPS,gross-preprint2017}.
We describe the algorithm and present the results below.
The code is available at \url{https://github.com/mmdanziger/disjointsetsxi}.
\begin{figure}
	\centering
	\begin{tikzpicture}[      
	every node/.style={anchor=north east,inner sep=0pt},
	x=1mm, y=1mm,
	]   
	\node (fig1) at (0,0)
	{\includegraphics[scale=0.7]{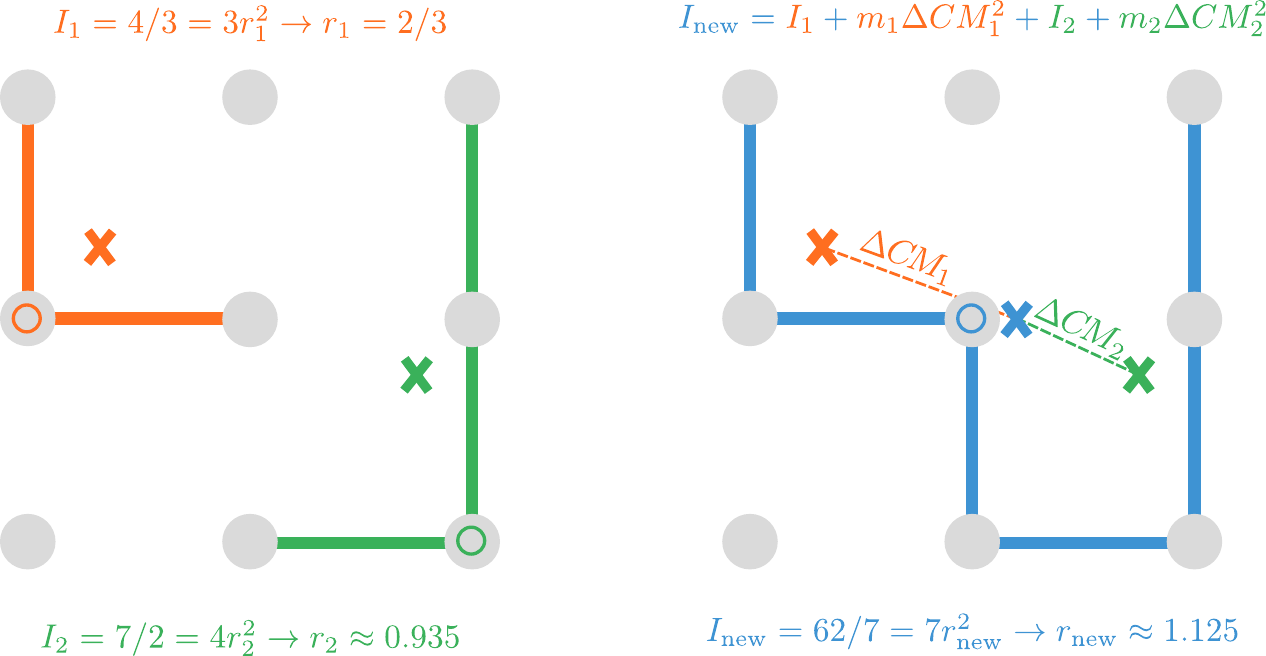}};
	
	\end{tikzpicture}
	\caption{\textbf{Tracking inertia and radius of gyration as links are added.} In the left panel, there are two clusters, with known inertia, and centers of mass marked with an "x".  In the right panel, a new link is added which joins the two clusters.  Instead of calculating the inertia from scratch, it can be updated using the parallel axis theorem as shown.  In this way, as links (or nodes) are added to the network we can continually track the inertia, the radius of gyration, and by averaging the radius, the correlation length $\xi$.}
	\label{fig:algorithm_diagram} 
	
\end{figure}

\section{Description}
We begin by assigning all nodes a unique cluster number.
We then add links (or sets of links corresponding to a node) one by one.
Each added link will either connect two nodes that are already in the same cluster and thereby have no effect on the connected components, or it will connect nodes from different clusters.
In such a case, we update the cluster number of the smaller cluster to point to the root of the larger cluster.
This is the standard case of disjoint sets, sometimes known as the Newman-Ziff percolation algorithm.

In order to measure the correlation length, we calculate the radius of gyration and moment of inertia of each cluster.
The moment of inertia of cluster $\mu$ is defined as:
\begin{equation}
I_\mu = \sum_{i}^{m_\mu} = (x_i - \bar{x}_{\mu})^2
\end{equation}
where  $i$ is the node index and $\bar{x}_{\mu}$ is the center of mass of the cluster.
The overall correlation length of the system is then defined as 
\begin{equation}\label{eq:xisum}
\xi^2 = \frac{\sum_{\mu} m_\mu I_\mu}{\sum_\mu m_\mu^2}.
\end{equation}
By the parallel axis theorem, when cluster 1 and cluster 2 are merged, the new cluster's moment of inertia will be:
\begin{equation}
I_{\rm new} = I_1 + m_1(\bar{x}_{\rm new} - \bar{x}_1)^2 + I_2 + m_2(\bar{x}_{\rm new} - \bar{x}_2)^2
\end{equation}
where $\bar{x}_{\rm new}$ is the new cluster's center of mass which is determined by the usual method:
\begin{equation}\label{eq:xbar_new}
\bar{x}_{\rm new} = \frac{m_1 \bar{x}_{1} + m_2 \bar{x}_{2}}{m_1 + m_2}.
\end{equation}

In the absence of periodic boundary conditions, this is sufficient to calculate $\xi$.
The introduction of periodic boundary conditions requires an additional step to insure that the node coordinates used to calculate the distances correspond to a consistent coordinate system.
For equation \ref{eq:xbar_new} to hold, $\bar{x}_{1}$ and $\bar{x}_{2}$ need to be measured from the same origin and so, when the clusters are merged we need to change the origin of the smaller cluster (2) to the origin of the larger cluster (1).
We assume, without loss of generality, that the clusters are numbered such that cluster 1 is larger than cluster 2.
We do this by examining the vector describing the merging link as measured in the coordinate systems of each cluster:
\begin{equation}
 dl = \tilde{x}_2 - \tilde{x}_1
\end{equation}
where $\tilde{x}_2$ ($\tilde{x}_1$) is the coordinate vector of the node in cluster 2 (1) where the link merging the clusters is connected.
We can then transform the center of mass of cluster 2 in the units counting from the origin of cluster 1 by
\begin{equation}
\bar{x}_{2}' = \bar{x}_{2} + \underbrace{\tilde{x}_1 + dl - \tilde{x}_2}_{=C}
\end{equation}
where $\bar{x}_{2}'$ is the coordinates of the center of mass of cluster 2, measured from the origin of cluster 1.
Without the effect of periodic boundary conditions, $C=0$ always.
Because the relative coordinates of each node are required (via $\tilde{x}_i$), all of the nodes in the smaller cluster need to be corrected with term $C$.


If a link connects back to the same component but $dl \neq \tilde{x}_2 - \tilde{x}_1$, then we have encountered a spanning component which should be removed from the calculation of $\xi$.
This also indicates that the percolation threshold has been reached.
Illustration of the algorithm can be seen in Fig. \ref{fig:algorithm_diagram}.

With this approach we can calculate $\xi$ above and below the percolation transition in a single sweep.
Above the transition, we simply ignore the spanning component in Eq. \ref{eq:xisum}.

\begin{figure}
	\centering
	\begin{tikzpicture}[      
	every node/.style={anchor=north east,inner sep=0pt},
	x=1mm, y=1mm,
	]   
	\node (fig1) at (0,0)
	{\includegraphics[scale=0.44]{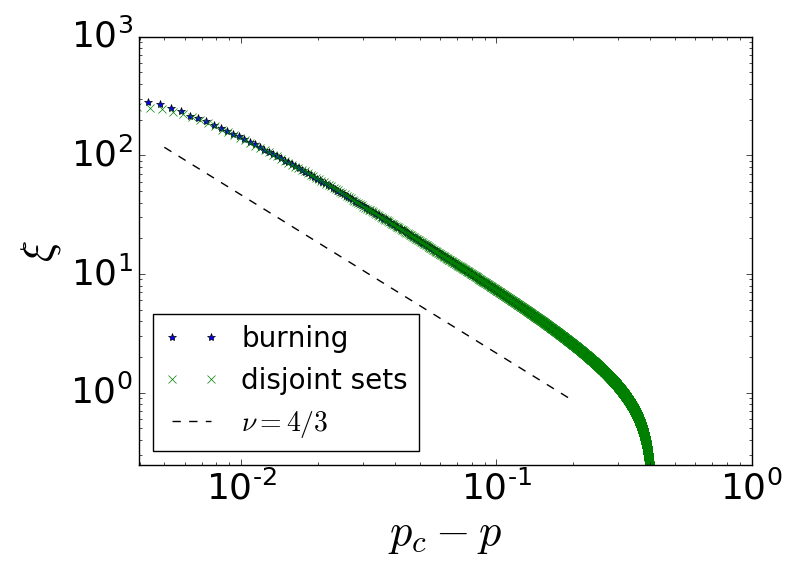}};
	\node (fig2) at (-7,-6)
	{\includegraphics[trim = 0 0 0 0 , scale=0.245,clip=true]{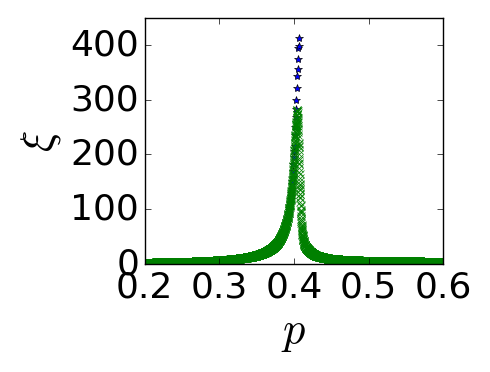}}; 
	\end{tikzpicture}
	\caption{\textbf{The percolation correlation length $\xi$ as a function of $p_c-p$ below criticality ($p<p_c$).} It is seen that both the burning and disjoint sets algorithms give similar results as $\xi \sim (p_c-p)^{-\nu}$ and $\nu=4/3$. The inset shows $\xi$ as function of $p$ without the scaling. Simulations shown for $N=10^6$ averaged over 1000 realizations. }
	\label{fig:xi_scaled} 
	
\end{figure}
\begin{figure}
	\centering
	\begin{tikzpicture}[      
	every node/.style={anchor=north east,inner sep=0pt},
	x=1mm, y=1mm,
	]   
	\node (fig1) at (0,0)
	{\includegraphics[scale=0.42]{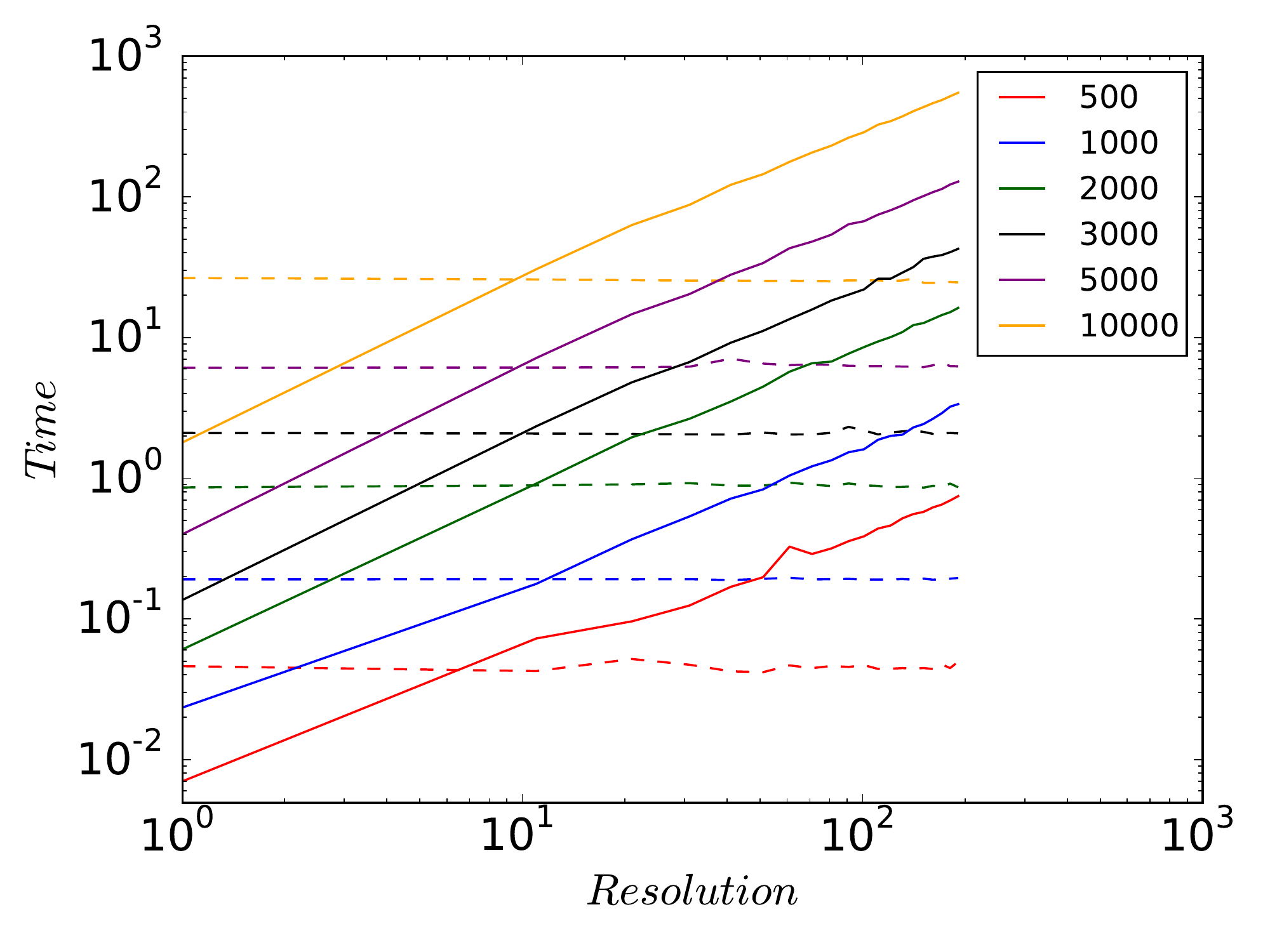}};
	
	\end{tikzpicture}
	\caption{\textbf{Running time as function of the resolution.} The continuous lines represent the computational running time of the burning algorithm for a given system size, and the dashed lines represent the running time of the new disjoint sets algorithm for the same system size. One can see that for resolution higher than $10$, the new disjoint sets algorithm is faster and the running time remains constant constant while the burning algorithm running time increases polynomially.}
	\label{fig:timevsresloglog} 
	
\end{figure}

\section{Results}
We compare the running time of the new algorithm discussed above against the classical one.
In Fig. \ref{fig:timevsresloglog} we show the computational running time of the algorithm as a function of the resolution of the measured quantity. In our case, the measurement for $\xi$  is done during percolation process from $p=0$ to $p=1$ with steps of $\Delta p$, and the resolution defined as $1/\Delta p$. The results for both algorithms below criticality can be seen in Fig. \ref{fig:xi_scaled} .\\
The continuous lines in Fig. \ref{fig:timevsresloglog} represent the running time of the classical algorithm and the dashed horizontal lines represent the running time of the new algorithm.\\
As one can see, for resolution higher than $10$ the new algorithm yields faster results compared to the cluster scanning one. 
Furthermore, the running time of the new algorithm stays constant no matter what the resolution is, while for the classic algorithm the running time increases linearly as the resolution increases.

\begin{figure}
 \includegraphics[width=\linewidth]{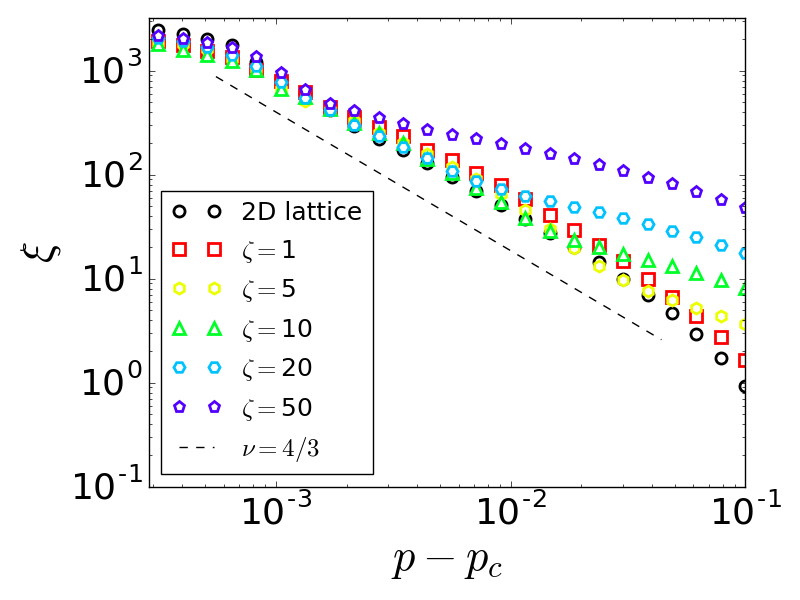}
 \caption{Measurement of the correlation length in a spatial network with exponential link lengths.
 In contrast to previous optimizations, the algorithm presented here can efficiently measure the correlation length in networks with spatial disorder.
 The network model was introduced in \cite{danziger-epl2016} and further analyzed in \cite{grossvaknin-JPS,gross-preprint2017,vaknin-njp2017}. \label{fig:xi_zeta}}
\end{figure}

In contrast to previous algorithms, we only need to assume that the nodes have well-defined positions in two-dimensional space and can calculate the correlation length for any link topology.
We demonstrate the utility of this by calculating the correlation length of the spatial topology introduced in \cite{danziger-epl2016}.  Here the nodes are assigned to a grid and the links are distributed randomly with geometric lengths drawn from an exponential distribution with average $\zeta$.
As we show in Fig. \ref{fig:xi_zeta} and study in depth in \cite{gross-preprint2017}, we are able to calculate the correlation length for different spatial topologies and show how their scaling regions differ as $\zeta$ is changed.

\acknowledgments{}

\bibliographystyle{apsrev4-1}

\bibliography{NoN.bib}
\end{document}